%% file: main.tex
\documentclass[10pt,conference]{IEEEtran}
\IEEEoverridecommandlockouts
\usepackage[utf8]{inputenc}
\usepackage[T1]{fontenc}
\usepackage{cite}
\usepackage{amsmath}
\usepackage{amssymb}
\usepackage{url}
\usepackage{amsfonts}
\usepackage{algpseudocode}
\usepackage{graphicx}
\usepackage{textcomp}
\usepackage{xcolor}
\usepackage{quantikz}
\usepackage{algorithm}
\usepackage{subcaption}
\usepackage{listings}
\begin{document}

\title{Gate Teleportation in Noisy Quantum Networks with the SquidASM Simulator}

\author{
    \IEEEauthorblockN{Valter Uotila}
    \IEEEauthorblockA{
    University of Helsinki\\
    valter.uotila@helsinki.fi}
}

\maketitle

\begin{abstract}
We implement the gate teleportation algorithm for teleporting arbitrary two-qubit Clifford gates and the Toffoli gate within the context of multi-node quantum networks, utilizing the SquidASM quantum network simulator. We show how a gate teleportation scheme can be used to implement gate cutting, which is an important approach to realize large circuits in distributed quantum computing environments. The correction operations in teleportation are automatically constructed for arbitrary two-qubit Clifford gates. We present simulation results for CNOT, DCNOT, CZ, SWAP, and Toffoli gates. For the Toffoli gate, we apply a similar gate teleportation protocol with the difference that the correction operation becomes more complex since the gate is non-Clifford. We perform the simulations under varying conditions of quantum channel and device noise levels. The simulations provide valuable insights into the robustness and efficacy of the implemented algorithms, and they assist in identifying the critical components within quantum networks where noise primarily affects the execution of applications.
\end{abstract}


\input{introduction}
\input{preliminaries}
\input{two_node}
\input{three_node}
\input{toffoli_teleportation}
\input{discussion}

\section*{Acknowledgment}
The author thanks Jukka K. Nurminen, Arianne Meijer–van de Griend, and Ilmo Salmenperä for helpful conversations and Quantum Internet Alliance (QIA) for selecting this work as one of the top three submissions for the first global Quantum Internet Application Challenge 2023.

\bibliographystyle{IEEEtran}
\bibliography{IEEEabrv, ref}

\end{document}

%% file: introduction.tex
\section{Introduction}

According to IBM's Quantum roadmap \cite{IBMQuantumRoadmap}, some of the near-future quantum computers will be constructed by interconnecting smaller quantum computers via quantum networks. These networks facilitate the connection of multiple smaller quantum computers and employ quantum and classical links to connect these quantum systems. As the number of qubits and their quality increases, these composite systems will allow us to execute larger quantum algorithms. However, orchestrating such quantum computing infrastructures for executing quantum-classical algorithms and software presents new challenges. 

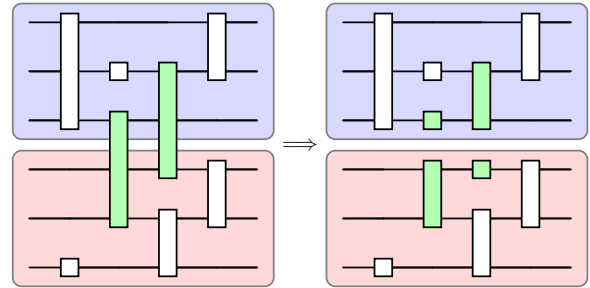
\begin{figure}[tbp]
    \centering
    \resizebox{0.9\columnwidth}{!}{
    \input{circuits/circuit_knitting_example}
    }
    \caption{Example demonstrating gate cutting between two quantum devices (blue and red)}
    \label{fig:circuit_knitting_example}
\end{figure}

An important collection of methods to execute larger circuits in distributed quantum computing environments are the circuit knitting techniques \cite{Piveteau_Sutter_2024}. The circuit knitting techniques can be divided into gate and wire cuttings \cite{gate_cutting}. Fig.~\ref{fig:circuit_knitting_example} demonstrates the conceptual idea behind gate cutting, which is a key application of this work. The standard gate-cutting procedure can be divided into three steps \cite{gate_cutting}. First, the circuits are decomposed into smaller circuits, which are visualized in Fig.~\ref{fig:circuit_knitting_example}). Then, the decomposed circuits are executed separately. Finally, the expectation value of the full-sized circuit can be reconstructed based on these subexperiments. The most efficient circuit knitting methods utilize quantum teleportation techniques to distribute the circuit on multiple quantum computers across the infrastructure. By harnessing quantum teleportation, the classical overhead in circuit knitting techniques can be significantly reduced, paving the way for more scalable and efficient quantum computations.

Given the central role of quantum teleportation in the development of future quantum-centric supercomputers, our research considers the practical implementation of quantum gate teleportation protocols within noisy quantum network environments. This study focuses on demonstrating, at the software engineering level, the feasibility of gate teleportation in noisy multi-node quantum networks using the SquidASM quantum network simulator \cite{squidasm}. SquidASM is built on top of NetSquid \cite{Coopmans_2021}, which utilizes NetQASM \cite{Dahlberg_2022} as a low-level instruction set architecture for hybrid quantum-classical programs in a quantum network or quantum internet. It allows simulations and control over various noise models for quantum devices, quantum channels, and classical channels.

Quantum teleportation research has been mainly focused on experimental research \cite{Pirandola_Eisert_Weedbrook_Furusawa_Braunstein_2015, Hu_Guo_Liu_Li_Guo_2023} on realizing quantum teleportation in various forms on different quantum computing devices. The standard and well-known algorithms to perform the state and gate teleportations are presented in \cite{Nielsen_Chuang_2010, Gottesman_Chuang_1999}. These ideas are widely applied in quantum error correction and the development of fault-tolerant quantum computers. Besides them, the algorithmic application of quantum networks is the quantum key-distribution-related protocols \cite{Bennett_Brassard_2014}, blind quantum computing \cite{5438603}, and quantum walks in distributed quantum computing \cite{9605308}. 




The key contributions of this paper are outlined as follows:
\begin{itemize}
    \item We develop and implement a general algorithm to calculate the correction operator for arbitrary two-qubit Clifford gates.
    \item We implement noisy and noiseless two- and three-node quantum network simulations for teleporting an arbitrary two-qubit state through CNOT, double CNOT, controlled-Z, and SWAP gates.
    \item We develop and implement Toffoli gate teleportation in a four-node quantum network. Teleportation of non-Clifford gates is not often demonstrated in the literature.
    \item We perform comprehensive simulations on noisy and noiseless quantum devices and quantum channels for the gate teleportation of the selected gates. Our implementation \cite{SquidKnit} is on GitHub.
\end{itemize}


We start by introducing standard state teleportation and its connection to gate teleportation and briefly describe the SquidASM simulator. Then, we simulate two-qubit Clifford gate teleportation in two-node quantum networks with noisy quantum devices and channels. This work naturally extends to quantum networks consisting of three or more nodes. We then develop and describe the implementation of Toffoli gate teleportation which admits more complex correction operation. For each implementation, we present simulation results and discuss their implications. Finally, we conclude the paper and discuss future work.

%% file: circuits/circuit_knitting_example.tex
\begin{quantikz}
 \qw \gategroup[3,steps=6,style={rounded corners,fill=blue!30,opacity=0.5,inner ysep=1pt}, background]{} & \gate[3]{} & \qw & \qw & \gate[2]{} & \qw \\
 \qw & \qw & \gate[1]{} & \gate[3, style={fill=green!30}]{} & \qw & \qw \\
 \qw & \qw & \gate[3, style={fill=green!30}]{} & \qw & \qw & \qw \\
 \qw \gategroup[3,steps=6,style={rounded corners,fill=red!30,opacity=0.5,inner ysep=1pt}, background]{} & \qw & \qw & \qw & \gate[2]{} & \qw \\
 \qw & \qw & \qw & \gate[2]{} & \qw & \qw \\
 \qw & \gate[1]{} & \qw & \qw & \qw & \qw
\end{quantikz} $\Longrightarrow$
\begin{quantikz}
 \qw \gategroup[3,steps=6,style={rounded corners,fill=blue!30,opacity=0.5,inner ysep=1pt}, background]{} & \gate[3]{} & \qw & \qw & \gate[2]{} & \qw \\
 \qw & \qw & \gate[1]{} & \gate[2, style={fill=green!30}]{} & \qw & \qw \\
 \qw & \qw & \gate[1, style={fill=green!30}]{} & \qw & \qw & \qw \\
 \qw \gategroup[3,steps=6,style={rounded corners,fill=red!30,opacity=0.5,inner ysep=1pt}, background]{} & \qw & \gate[2, style={fill=green!30}]{} & \gate[1, style={fill=green!30}]{} \qw & \gate[2]{} & \qw \\
 \qw & \qw & \qw & \gate[2]{} & \qw & \qw \\
 \qw & \gate[1]{} & \qw & \qw & \qw & \qw
\end{quantikz}


%% file: preliminaries.tex
\section{Preliminaries}

\subsection{State teleportation}
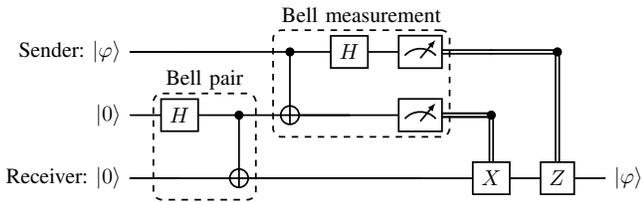
\begin{figure}[tbp]
    \centering
    \resizebox{\columnwidth}{!}{%
    \input{circuits/quantum_teleportation}
    }
    \caption{Standard circuit to implement quantum state teleportation}
    \label{fig:quantum_teleportation}
\end{figure}

Quantum state teleportation is a standard quantum information theoretical technique to transfer an exact state between two parties using an entangled pair and classical communication \cite{Gottesman_Chuang_1999, Nielsen_Chuang_2010}. We briefly outline its idea to establish a connection with gate teleportation. Usually, the state teleportation scheme is described with the circuit diagram in Fig. \ref{fig:quantum_teleportation}.

Initially, the sender has a state $| \varphi \rangle$, which they want to transfer to the receiver. The sender and receiver share an entangled pair (Bell pair), which can be realized over a quantum channel. After initializing the entangled pair, the receiver's state is in one of the four states: $|\varphi\rangle$, $X|\varphi\rangle$, $Z|\varphi\rangle$, $XZ|\varphi\rangle$. Depending on the bits the sender measures in their Bell measurement, the receiver performs a correction operation by applying $X$ and $Z$ gates to their qubit. The information about which corrections the receiver should perform is transferred over a classical channel.

\begin{figure}[tbp]
    \centering
    \includegraphics[width = 0.9\columnwidth]{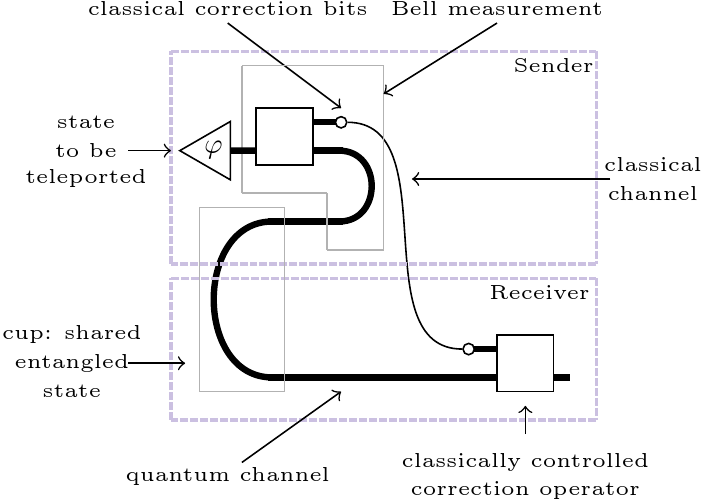}
    \caption{State teleportation utilizing the diagrammatic notation from categorical quantum mechanics}
    \label{fig:teleportation_diagram1}
\end{figure}

An alternative method to understand quantum teleportation is to use the notation that Penrose introduced \cite{penrose1971applications} and which has been further developed in categorical quantum mechanics \cite{Coecke_2010, Coecke_Kissinger_2017}. We use it to describe how we can construct the circuits to perform gate teleportation. With the notation, quantum teleportation has the diagrammatic expression shown in Fig.~\ref{fig:teleportation_diagram1}. The so-called ''cups'' are mapped to Bell states, and the ''caps'' are mapped to Bell measurements. The classical information is transformed in a classical wire connecting the Bell measurement and the correction operator. The explanation of why the diagrammatic reasoning is formally valid is represented in \cite{Coecke_Kissinger_2017}.

\subsection{Gate teleportation}

State teleportation aims to transfer a state between two parties without modifying it. Compared to state teleportation, gate teleportation extends the approach by applying a given gate to the teleported state. Alternatively, it is commonly described as teleporting the state through a gate. Gate teleportation was introduced in \cite{Gottesman_Chuang_1999, Gottesman_Chuang_1999_universal} as a universal primitive to perform fault-tolerant quantum computations. One of the key results of \cite{Gottesman_Chuang_1999} is the fact that the work introduces a universal gate set that does not require two-qubit gates except measurement.

\begin{figure}[tbp]
    \centering
    \resizebox{\columnwidth}{!}{%
    \input{circuits/single_gate_teleportation}
    }
    \caption{Circuit to implement gate teleportation through gate $U$ with the correction operation $R_{xy}$ where $x$ and $y$ depend on the Bell measurement result. The figure is similar to Fig.~\ref{fig:quantum_teleportation} except of application $U$ and the $U$-dependent correction operation $R_{xy}$.}
    \label{fig:single_gate_teleportation}
\end{figure}
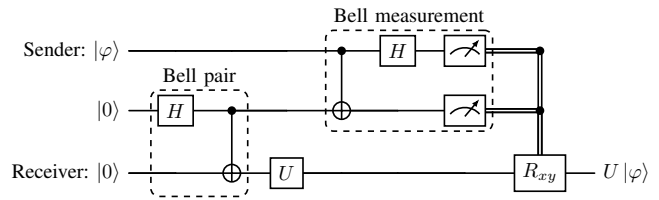

Fig.~\ref{fig:single_gate_teleportation} demonstrates the standard gate teleportation for a single-qubit unitary $U$. The interesting fact that we can read from the circuit is that we can apply the gate $U$ before the state $|\varphi\rangle$ is even prepared. Additionally, all the operations before the gate $U$ are only Hadamard and CNOT, which do not depend on the state $| \varphi \rangle$.

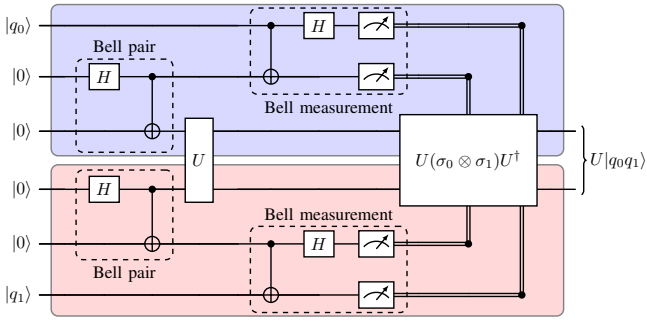
\begin{figure}[tbp]
    \centering
    \resizebox{\columnwidth}{!}{%
    \input{circuits/two_machines_gate_teleportation}
    }
    \caption{Circuit to implement gate cutting with gate teleportation based on Fig. 5 in \cite{Piveteau_Sutter_2024}. The gate $U$ is performed between blue and red devices.}
    \label{fig:two_machines_gate_teleportation}
\end{figure}

The single-qubit gate teleportation scheme generalizes to two-qubit gates and $n$-qubit gates. The circuit in Fig.~\ref{fig:two_machines_gate_teleportation} describes the algorithm of teleporting two-qubit Clifford gate $U$. In this work, we focus on the two-qubit Clifford gates and the Toffoli gate. These two cases differ in the following sense \cite{Gottesman_Chuang_1999_universal}: Let $C_1$ be the Pauli group $\langle X, Y, Z \rangle$ which is generated by the Pauli matrices $X$, $Y$ and $Z$. Let $C_2$ be the Clifford group, which is the set of unitaries that map Pauli gates into Pauli gates:
\begin{equation*}
    C_2 := \left\{ U \mid U \sigma U^{\dag} \in C_1 \text{ for all } \sigma \in C_1 \right\}.
\end{equation*}
Now, CNOT, DCNOT, SWAP, and CZ gates belong to $C_2$ \cite{Gottesman_Chuang_1999_universal}. On the other hand, we can continue this construction recursively and define
\begin{equation*}
    C_3 := \left\{ U \mid U \sigma U^{\dag} \in C_2 \text{ for all } \sigma \in C_1 \right\}.
\end{equation*}
Generally, we can define
\begin{equation}\label{eq:clifford_k}
    C_k := \left\{ U \mid U \sigma U^{\dag} \in C_{k-1} \text{ for all } \sigma \in C_1 \right\}.
\end{equation}

We know that the Toffoli gate belongs to $C_3$ \cite{Gottesman_Chuang_1999_universal}. Depending on whether the gate belongs to set $C_2$ or $C_3$, it requires a distinct type of correction operation (e.g., $R_{xy}$ in Fig.~\ref{fig:single_gate_teleportation}). In this work, we will concretely construct these correction operations for the selected two-qubit gates and the Toffoli gate.

Next, we describe the method to construct a correction operation for a two-qubit Clifford gate $U$. A similar approach applies to the Toffoli gate. After the sender's two-qubit state has been teleported through the gate $U$, the receiver is left with four options: $U(X \otimes I)|\varphi\rangle$, $U(Z \otimes I)|\varphi\rangle$, $U(I \otimes X)|\varphi\rangle$, $U(I \otimes Z)|\varphi\rangle$. Since we want to output the state $U|\varphi\rangle$, we can rewrite these states as 
\begin{align}\label{eq:rewritten_gate_teleportation_states}
    U(X \otimes I)|\varphi\rangle =& \ (U(X \otimes I)U^{\dag})U|\varphi\rangle, \notag\\
    U(Z \otimes I)|\varphi\rangle =& \ (U(Z \otimes I)U^{\dag})U|\varphi\rangle,\\
    U(I \otimes X)|\varphi\rangle =& \ (U(I \otimes X)U^{\dag})U|\varphi\rangle \text{ and } \notag\\
    U(I \otimes Z)|\varphi\rangle =& \ (U(I \otimes Z)U^{\dag})U|\varphi\rangle. \notag
\end{align}
From these equations we can deduce that we need to find a correction operation which cancels effect of a gate of type $U(\sigma_0 \otimes \sigma_1)U^{\dag}$, where $\sigma_0, \sigma_1 \in \left\{I, Z, X\right\}$, that leaves us with the wanted result $U|\varphi\rangle$.

As pointed out in \cite{Gottesman_1998}, we want to study how the teleported gate $U$ transforms the basic operations $X \otimes I$, $Z \otimes I$, $I \otimes X$, and $I \otimes Z$. In essence, our objective is to determine the specific elements of the Pauli group to which the operator $U$ maps these basic elements. In order to realize the correction operation, this means that we calculate the Pauli decompositions for elements $U(I \otimes X)U^{\dag}$, $U(I \otimes Z)U^{\dag}$, $U(X \otimes I)U^{\dag}$ and $U(X \otimes I)U^{\dag}$ in \eqref{eq:rewritten_gate_teleportation_states}. The decompositions give the Pauli operations that are applied depending on the classical bits we obtain from the Bell measurements. The computation of the correction operation is represented in Alg. \ref{alg:correction_operator}. The algorithm uses a subroutine that decomposes the element of type $U(\sigma_0 \otimes \sigma_1)U^{\dag}$ into a linear combination of Pauli operators. The Pauli decomposition routine is implemented, for example, in Pennylane \cite{bergholm2022pennylane}.

\begin{figure*}[tbp]
    \centering
    \includegraphics[width=0.99\textwidth]{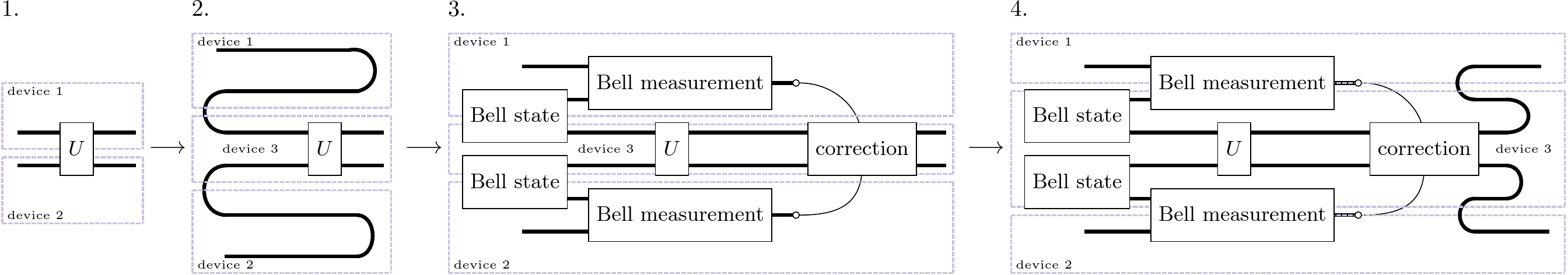}
    \caption{Diagram representing how to construct gate teleportation using the notation from categorical quantum mechanics, which is summarized in Fig.~\ref{fig:teleportation_diagram1}. Gate $U$ is distributed across devices 1 and 2, which is a concrete example of gate knitting presented in Fig.~\ref{fig:circuit_knitting_example}. The circuit diagram at step 3. has a concrete circuit represented in Fig.~\ref{fig:three_machines}.}
    \label{fig:diagram}
\end{figure*}

A simple diagrammatic description to construct an abstract gate teleportation protocol for two-qubit gates is represented in Fig. \ref{fig:diagram}. The intuitive idea behind the diagrammatic notation is that the wires in the system can be bent, which moves them into a different quantum device using a quantum channel in the quantum network. It is a relatively simple method to design algorithms. This is the main reason why the diagrammatic notation seems to be useful and natural in quantum networking. After bending, the wires have ''cups'' and ''caps'' which can be implemented with Bell states and Bell measurements with classically controlled correction operation at the end of the process. 

Fig. \ref{fig:diagram} also illustrates the concept of circuit knitting, focusing particularly on gate cutting \cite{gate_cutting}. In this scenario, the gate $U$ is distributed across device 1 and device 2, which means that one of the qubits resides on device 1 and the other qubit is on device 2. We need to perform a gate-cutting procedure for executing the circuit. Various methods exist for this purpose that include exact and approximate techniques \cite{gate_cutting}. This study introduces an alternative approach employing gate teleportation and a third quantum device to achieve precise gate cutting without classical overhead.

More precisely, at step $1.$ in Fig. \ref{fig:diagram}, we assume to have a gate $U$, which we want to execute across two devices. Instead of performing the circuit knitting (i.e., gate cutting) process described in \cite{Piveteau_Sutter_2024}, we introduce a third quantum device that we use to execute the gate $U$ precisely. The same motivation applies here as in the single-gate teleportation scheme: we only need to apply simple operations before the gate $U$ on the third machine. In real-life scenarios, we might not have resources left on devices 1 and 2, so we can utilize the teleporting scheme to execute the gate $U$ utilizing the available third quantum device. Now, at step $2.$, we bend the wires, which enables us to move the states from devices 1 and 2 to device 3. At step $3.$, we rewrite the ''cups'' and ''caps'' following their definitions \cite{Coecke_Kissinger_2017}. Finally, at step $4.$, we might want to consider teleporting the state back to devices 1 and 2 in order to continue the computation as it was left at step $1.$ This can be achieved by bending the wires back to the devices 1 and 2. Additionally, this makes the third device again available for further quantum computations.

\input{algorithms/correction_algorithm}

\subsection{Quantum networks and SquidASM simulator}

Quantum networks \cite{Simon_2017} form an important part of quantum computing, quantum information processing, and quantum communications. Quantum networks are designed to transmit qubits between physically separated quantum devices, which are connected with quantum channels and classical channels.

\begin{figure}[tbp]
    \centering
    \includegraphics[width = \columnwidth]{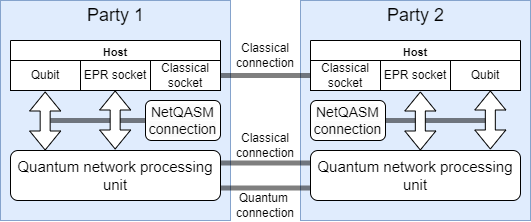}
    \caption{SquidASM's program context \cite{squidasm}}
    \label{fig:squidasm_program_context}
\end{figure}

SquidASM \cite{squidasm} is a quantum network simulator based on NetSquid \cite{Coopmans_2021} that can execute applications written using NetQASM \cite{Dahlberg_2022}. NetSquid is a Network Simulator for Quantum Information using Discrete events. NetQASM serves as a low-level instruction set architecture designed to facilitate interaction with quantum network controllers and to execute applications within a quantum network. 
Other quantum network simulators are QuISP \cite{QuISP}, SimulaQron \cite{SimulaQron}, QuNetSim \cite{QuNetSim}, and SeQUeNCe \cite{SeQUeNCe}. Our implementation \cite{SquidKnit} on GitHub follows the recommended SquidASM program structure having \texttt{application.py} and \texttt{run.py} files for both two-node and three-node simulations. The \texttt{application.py} files contain the main logic, which is described in this paper. Fig.~\ref{fig:squidasm_program_context} describes SquidASM's program context in the case of a two-node quantum network.

%% file: circuits/quantum_teleportation.tex
\begin{quantikz}
  \lstick{Sender: $\ket{\varphi}$} & \qw & \qw & \ctrl{1} \gategroup[2,steps=3,style={dashed, rounded corners, inner ysep=-1pt, inner xsep=0pt}]{Bell measurement} & \gate{H} & \meter{} & \setwiretype{c} & \ctrl[vertical wire=c]{2} \\
  \lstick{$\ket{0}$} & \gate{H} \gategroup[2,steps=2,style={dashed, rounded corners, inner ysep=-1pt, inner xsep=0pt}]{Bell pair} & \ctrl{1} & \targ{}  & \qw & \meter{} & \cwbend{1} \setwiretype{c} \\
  \lstick{Receiver: $\ket{0}$} & \qw & \targ{} & \qw & \qw & \qw & \gate{X} & \gate{Z} & \rstick{$\ket{\varphi}$}
\end{quantikz}

%% file: circuits/single_gate_teleportation.tex
\begin{quantikz}
  \lstick{Sender: $\ket{\varphi}$} & \qw & \qw & \qw & \ctrl{1} \gategroup[2,steps=3,style={dashed, rounded corners, inner ysep=-1pt, inner xsep=0pt}]{Bell measurement} & \gate{H} & \meter{} & \cwbend{2} \setwiretype{c} \\
  \lstick{$\ket{0}$} & \gate{H} \gategroup[2,steps=2,style={dashed, rounded corners, inner ysep=-1pt, inner xsep=0pt}]{Bell pair} & \ctrl{1} & \qw & \targ{}  & \qw & \meter{} & \cwbend{1} \setwiretype{c} \\
  \lstick{Receiver: $\ket{0}$} & \qw & \targ{} & \gate{U} & \qw & \qw & \qw & \gate{R_{xy}} & \rstick{$U\ket{\varphi}$}
\end{quantikz}

%% file: circuits/two_machines_gate_teleportation.tex
\begin{quantikz}
\lstick{$|q_0\rangle$} & \qw \gategroup[3,steps=12,style={rounded corners,fill=blue!30,opacity=0.5,inner ysep=1pt}, background]{}   & \qw      & \qw      & \qw & \qw          & \ctrl{1}\gategroup[2,steps=3,style={dashed,rounded corners, inner ysep=-1pt}, label style={label
position=below,anchor=north,yshift=-0.2cm}]{Bell measurement} & \gate{H} & \meter{} & \setwiretype{c} & & \ctrl[vertical wire=c]{2} & \wireoverride{n} \\
\lstick{\ket{0}} & \qw & \gate{H}\gategroup[2,steps=2,style={dashed,rounded corners, inner xsep=4pt, inner ysep=-1pt}]{Bell pair} & \ctrl{1} & \qw & \qw & \targ    & \qw      & \qw      & \meter{} & \setwiretype{c} & \ctrl[vertical wire=c]{1} & \wireoverride{n} \\
\lstick{\ket{0}} & \qw & \qw      & \targ    & \qw & \gate[2]{U} & \qw & \qw      & \qw           & \qw & \qw & \gate[2,style={inner xsep=25pt}]{\hspace*{-20pt}U (\sigma_0 \otimes \sigma_ 1) U^{\dag}\hspace*{-20pt}} & \qw & \qw & \qw \rstick[2]{$U|q_0q_1\rangle$} \\
\lstick{\ket{0}} & \qw \gategroup[3,steps=12,style={rounded corners,fill=red!30,opacity=0.5,inner ysep=1pt}, background]{} & \gate{H}\gategroup[2,steps=2,style={dashed,rounded corners, inner xsep=4pt, inner ysep=-2pt},label style={label
position=below,anchor=north,yshift=-0.2cm}]{Bell pair} & \ctrl{1} & \qw &         & \qw & \qw      & \qw      & \qw      & \qw  & & & \\
\lstick{\ket{0}} & \qw & \qw      & \targ    & \qw           & \qw & \qw      & \ctrl{1}\gategroup[2,steps=3,style={dashed,rounded corners, inner ysep=-1pt}]{Bell measurement} & \gate{H} & \meter{} & \setwiretype{c} & \ctrl[vertical wire=c]{-1} & \wireoverride{n} \\
\lstick{$|q_1\rangle$} & \qw & \qw & \qw & \qw & \qw  & \targ & \qw & \qw & \meter{} & \setwiretype{c} & & \ctrl[vertical wire=c]{-2} & \wireoverride{n}
\end{quantikz}

%% file: algorithms/correction_algorithm.tex
\begin{algorithm}[htpb]
\caption{Construction of the correction operator}
\label{alg:correction_operator}
\begin{algorithmic}[1]
\State \textbf{Input:} Two-qubit Clifford gate $U$
\State \textbf{Output:} correction\_operators
\State Initialize correction\_operator as an empty list
\State basis $\gets$ [$X \otimes I$, $Z \otimes I$, $I \otimes X$, $I \otimes Z$]
\For{each $V$ in basis}
    \State op $\gets$ $UVU^{\dag}$
    \State $\sigma_0 \otimes \sigma_1$ $\gets$ pauli\_decompose(op)
    \State append $\sigma_0 \otimes \sigma_1$ to correction\_operator
\EndFor
\State \Return correction\_operator
\end{algorithmic}
\end{algorithm}

%% file: two_node.tex
\section{Quantum network simulations}

\subsection{Two-node network simulations}

In this part, we describe how to implement a two-node quantum network simulation for the two-qubit Clifford gates. Precisely, we realize the circuit in Fig.~\ref{fig:two_machines_gate_teleportation} with SquidASM in a two-node quantum network. The circuit in Fig.~\ref{fig:two_machines_gate_teleportation} is based on the circuit in Fig. 5 in \cite{Piveteau_Sutter_2024}, and we reproduced it with the diagrammatical notation in Fig.~\ref{fig:diagram}. The original reasoning why this circuit performs gate teleportation is also explained in \cite{Gottesman_Chuang_1999, Gottesman_Chuang_1999_universal}.

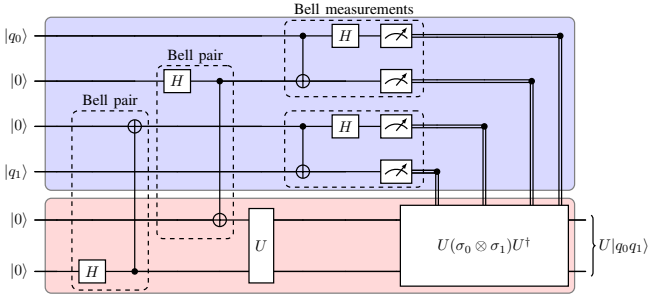
\begin{figure}[tbp]
    \centering
    \resizebox{\columnwidth}{!}{%
    \input{circuits/two_machines_gate_teleportation_reorganized}
    }
    \caption{This figure reorganizes the qubits in Fig. \ref{fig:two_machines_gate_teleportation}. The construction is a two-qubit generalization of the circuit in Fig.~\ref{fig:single_gate_teleportation}.}
    \label{fig:two_machines_gate_teleportation_reorganized}
\end{figure}

The two-node quantum network is the simplest possible quantum network consisting of two nodes and one link that consists of a quantum channel and a classical channel. In order to realize the circuit in Fig.~\ref{fig:two_machines_gate_teleportation}, we reorganize the qubits so that the entangled pairs are between the quantum devices. The result of this reorganization is represented in Fig.~\ref{fig:two_machines_gate_teleportation_reorganized}. The first four qubits with the blue background are on the first machine, which contains the two-qubit input state $|q_0 q_1 \rangle$ to the gate $U$. The last two qubits on the red background are on the second machine, which outputs the result $U|q_0 q_1 \rangle$. The machines share two Bell pairs. The classical channel delivers four bits of information, which are used to correct the receiver's state using Alg.~\ref{alg:correction_operator}.

We omit the pseudocode for this case since it is closely similar to the three-node simulation which we describe next.

%% file: circuits/two_machines_gate_teleportation_reorganized.tex
\begin{quantikz}
\lstick{$|q_0\rangle$} & \qw\gategroup[4,steps=13,style={rounded corners,fill=blue!30,opacity=0.5,inner ysep=1pt}, background]{} & \qw & \qw & \qw & \qw & \qw & \ctrl{1} \gategroup[2,steps=3,style={dashed,rounded corners, inner ysep=-1pt}, label style={label
position=above,anchor=north,yshift=0.3cm}]{Bell measurements} & \gate{H} & \meter{} & \setwiretype{c} & & & \ctrl[vertical wire=c]{4} \\
\lstick{\ket{0}} & \qw & \qw & \qw & \gate{H}\gategroup[4,steps=2,style={dashed,rounded corners, inner xsep=1pt, inner ysep=-1pt}]{Bell pair} & \ctrl{3} & \qw & \targ & \qw & \qw & \meter{} & \setwiretype{c} & & \ctrl[vertical wire=c]{3} \\
\lstick{\ket{0}} & \qw & \qw\gategroup[4,steps=2,style={dashed,rounded corners, inner xsep=1pt, inner ysep=-1pt}]{Bell pair} & \targ  & \qw & \qw & \qw & \qw & \ctrl{1}\gategroup[2,steps=3,style={dashed,rounded corners, inner ysep=-1pt}]{} & \gate{H} & \meter{} & \setwiretype{c} & \ctrl[vertical wire=c]{2} \\
\lstick{$|q_1\rangle$} &\qw &  \qw & \qw & \qw& \qw & \qw & \targ & \qw& \qw& \meter{} \wire[r][1]{c} & \ctrl[vertical wire=c]{1} \setwiretype{c} \\
\lstick{\ket{0}} & \qw\gategroup[2,steps=13,style={rounded corners,fill=red!30,opacity=0.5,inner ysep=1pt}, background]{} & \qw& \qw & \qw & \targ & \qw & \gate[2]{U} & \qw& \qw & \qw & \qw & \gate[2,style={inner xsep=40pt}]{\hspace*{-20pt}U (\sigma_0 \otimes \sigma_ 1) U^{\dag}\hspace*{-20pt}} & \qw & \qw & \qw \rstick[2]{$U|q_0q_1\rangle$} \\
\lstick{\ket{0}} & \qw & \gate{H} & \ctrl{-3} & \qw& \qw & \qw & \qw& \qw& \qw & \qw & \qw & \qw& \qw& \qw
\end{quantikz}

%% file: three_node.tex
\subsection{Three-node network simulations} 

\begin{figure}[tbp]
    \centering
    \resizebox{\columnwidth}{!}{%
    \input{circuits/three_machines}
    }
    \caption{Circuit implementing the gate teleportation using three quantum devices, i.e., three nodes in the quantum network (blue, yellow, red). This circuit is a concrete instance of the circuit diagram in step 3. in Fig.~\ref{fig:diagram}.}
    \label{fig:three_machines}
\end{figure}
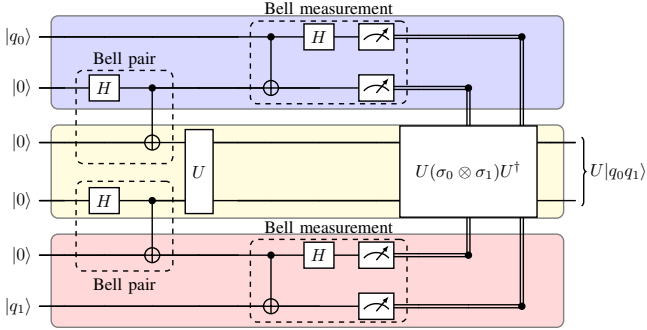

The three-node network simulation is closely similar to the two-node network simulation. We rewrite the diagrammatic construction in Fig. \ref{fig:diagram} as an actual circuit in Fig. \ref{fig:three_machines}, which works as a basis for the implementation. The blue and red parts are located on machines 1 and 2. The yellow part corresponds to the operations performed on machine 3.

The three-node quantum gate teleportation simulation proceeds by applying steps in Alg. \ref{alg:input_program} and Alg. \ref{alg:gate_program}. We name the first algorithm InputProgram since the input state is prepared there and the second algorithm GateProgram because the teleported gate $U$ is executed there. Since we have two input qubits residing on different quantum devices, we need to initialize two instances of InputProgram. These instances correspond to the devices 1 and 2 in Fig. \ref{fig:diagram}. Similarly, these instances correspond to the devices that implement the blue and red operations in Fig. \ref{fig:three_machines}. In practice, these programs would run independently on different quantum computers. GateProgram implements the logic on machine 3 in Fig. \ref{fig:diagram} or, similarly, the logic on the yellow machine in Fig. \ref{fig:three_machines}.

\input{algorithms/input_program}

\input{algorithms/gate_program}

Our overall goal is to apply the gate $U$ to two input qubits using the gate teleportation protocol. In practice, this is achieved by running InputProgram in Alg. \ref{alg:input_program} and GateProgram in Alg. \ref{alg:gate_program} simultaneously using SquidASM. The algorithms share the same context in SquidASM, which describes the quantum network. All three programs can get access to the classical channels (csockets) and the quantum channels (epr\_sockets).

InputPrograms prepare two initial qubits with the gates that the user provides in \texttt{gates} input parameter (lines 8-9 in Alg. \ref{alg:input_program}). These qubits form the state, which is the input for the gate $U$. The GateProgram receives the shared Bell pairs from the InputPrograms (lines 10-11 in Alg. \ref{alg:gate_program}). Then, the Bell measurement is prepared on both InputPrograms (lines 13-14 in Alg. \ref{alg:input_program}), and the four bits of information from the measurements are transferred with the classical channels to the GateProgram (lines 18-19 in Alg. \ref{alg:input_program}). As the pseudo-code for GateProgram in Alg. \ref{alg:gate_program} shows, the gate $U$ is not applied to the qubits in InputPrograms but to the qubits initialized from the quantum channel (line 12 in Alg. \ref{alg:gate_program}). 

The GateProgram receives four bits of classical information (lines 16-17 in Alg. \ref{alg:gate_program}) from the Bell measurements. Based on the classical bits and the precalculated correction operation, the required corrections are performed (lines 20-31 in Alg. \ref{alg:gate_program}). The correction operation is calculated with Alg.~\ref{alg:correction_operator}. The pseudo-code for two-node gate teleportation is similar, except that we have only a single InputProgram that implements the same logic that is now divided between the two programs.

%% file: circuits/three_machines.tex
\begin{quantikz}
\lstick{$|q_0\rangle$} & \qw \gategroup[2,steps=12,style={rounded corners,fill=blue!30,opacity=0.5,inner ysep=1pt}, background]{}   & \qw      & \qw      & \qw & \qw          & \ctrl{1}\gategroup[2,steps=3,style={dashed,rounded corners, inner ysep=-2pt}]{Bell measurement} & \gate{H} & \meter{} & \setwiretype{c} & & \ctrl[vertical wire=c]{2} & \wireoverride{n} \\
\lstick{\ket{0}} & \qw & \gate{H}\gategroup[2,steps=2,style={dashed,rounded corners, inner xsep=4pt, inner ysep=-1pt}]{Bell pair} & \ctrl{1} & \qw & \qw & \targ    & \qw      & \qw      & \meter{} & \setwiretype{c} & \ctrl[vertical wire=c]{1} & \wireoverride{n} \\
\lstick{\ket{0}} & \qw \gategroup[2,steps=12,style={rounded corners,fill=yellow!30,opacity=0.5,inner ysep=-3pt}, background]{} & \qw      & \targ    & \qw & \gate[2]{U} & \qw & \qw      & \qw           & \qw & \qw & \gate[2,style={inner xsep=25pt}]{\hspace*{-20pt}U (\sigma_0 \otimes \sigma_ 1) U^{\dag}\hspace*{-20pt}} & \qw & \qw & \qw \rstick[2]{$U|q_0q_1\rangle$} \\
\lstick{\ket{0}} & \qw & \gate{H}\gategroup[2,steps=2,style={dashed,rounded corners, inner xsep=4pt, inner ysep=-2pt},label style={label
position=below,anchor=north,yshift=-0.2cm}]{Bell pair} & \ctrl{1} & \qw &         & \qw & \qw      & \qw      & \qw      & \qw  & & & \\
\lstick{\ket{0}} & \qw \gategroup[2,steps=12,style={rounded corners,fill=red!30,opacity=0.5,inner ysep=1pt}, background]{} & \qw      & \targ    & \qw           & \qw & \qw      & \ctrl{1}\gategroup[2,steps=3,style={dashed,rounded corners, inner ysep=-2pt}]{Bell measurement} & \gate{H} & \meter{} & \setwiretype{c} & \ctrl[vertical wire=c]{-1} & \wireoverride{n} \\
\lstick{$|q_1\rangle$} & \qw & \qw & \qw & \qw & \qw  & \targ & \qw & \qw & \meter{} & \setwiretype{c} & & \ctrl[vertical wire=c]{-2} & \wireoverride{n}
\end{quantikz}

%% file: algorithms/input_program.tex
\begin{algorithm}[htpb]
\caption{Pseudocode for InputProgram}
\label{alg:input_program}
\begin{algorithmic}[1]
\Procedure{InputProgram}{context, gates}
\State Get access to EPR and classical channels from the program context
\State connection $\gets$ context.connection \Comment{Access global connection between the quantum devices}
\State Receive EPR qubit (epr) from gate device over the quantum channel
\State \textbf{yield from} connection.flush() \Comment{Wait for other devices by flushing the connection}
\State input qubit $\gets$ Create new qubit
\For{each gate in gates}
\State Apply gate to input qubit
\EndFor
\State \textbf{yield from} connection.flush() \Comment{Wait for other devices by flushing the connection}
\State Prepare and perform Bell measurement:
\State Apply CNOT between input and epr qubit
\State Apply Hadamard to input qubit
\State epr\_meas $\gets$ Measure epr qubit
\State input\_meas $\gets$ Measure input qubit
\State \textbf{yield from} connection.flush() \Comment{Wait for other devices by flushing the connection}
\State result $\gets$ (epr\_meas, input\_meas)
\State Send measurement results over the classical channel to the gate program to be used in the correction operation
\EndProcedure
\end{algorithmic}
\end{algorithm}

%% file: algorithms/gate_program.tex
\begin{algorithm}[htpb]
\caption{Pseudocode for GateProgram}
\label{alg:gate_program}
\begin{algorithmic}[1]
\Procedure{GateProgram}{context, gate $U$}
\State correction\_operator $\gets$ get\_correction\_operator($U$)
\State Get access to EPR and classical channels from the program context
\State connection $\gets$ context.connection \Comment{Access global connection between the quantum devices}
\State Receive EPR pairs (epr1, epr2) from input devices over the quantum channels
\State Apply gate $U$ to epr1 and epr2 qubits
\State \textbf{yield from} connection.flush() \Comment{Wait for input devices to finish Bell measurements by flushing the connection}
\State bits $\gets$ [bit0, bit1, bit2, bit3] \Comment{Receive four bits containing the measurement results over the classical channels}
\State meas\_to\_ops $\gets$ zip(bits, correction\_operator)
\For{each [meas, op] in meas\_to\_ops}
\State \textbf{if} meas = 1 \textbf{then}
\State \quad \textbf{if} op[0] = $\sigma_x$ \textbf{then}
\State \quad \quad Apply $X$ gate to epr2 qubit
\State \quad \textbf{elseif} op[0] = $\sigma_z$ \textbf{then}
\State \quad \quad Apply $Z$ gate to epr2 qubit
\State \quad \textbf{if} op[1] = $\sigma_x$ \textbf{then}
\State \quad \quad Apply $X$ gate to epr1 qubit
\State \quad \textbf{elseif} op[1] = $\sigma_z$ \textbf{then}
\State \quad \quad Apply $Z$ gate to epr1 qubit
\State \textbf{end if}
\EndFor
\EndProcedure
\end{algorithmic}
\end{algorithm}

%% file: toffoli_teleportation.tex
\section{Toffoli gate teleportation in four node quantum network}

In this section, we extend the previous approaches and teleport the non-Clifford Toffoli gate. When teleporting Clifford gates, we were able to calculate and realize the correction operation in terms of simple Pauli matrices. In the Clifford cases, the Pauli gates were always applied to a single qubit in order to correct the result. In the non-Clifford case, we are not able to find a similar local correction operation. On the other hand, we are not interested in limiting ourselves to operating only with Pauli matrices. Instead of using only Pauli gates, we will construct more advanced corrections operations expressed as Hamiltonians based on the Pauli decompositions of $\mathrm{TOFF}(\sigma_1 \otimes \sigma_2 \otimes \sigma_3)\mathrm{TOFF}$ where $\mathrm{TOFF}$ is the Toffoli gate and $\sigma_i$, for $i = 1,2,3$, are selected so that they form the basis as described previously in \eqref{eq:rewritten_gate_teleportation_states}. These Hamiltonians will provide us with a method to correct the Toffoli gate teleportation result.

Recalling the reasoning that earlier led to \eqref{eq:rewritten_gate_teleportation_states}, we calculate Pauli decompositions for the following matrices
\begin{align*}
    \mathrm{TOFF}(Z \otimes I \otimes I)\mathrm{TOFF},\ \mathrm{TOFF}(X \otimes I \otimes I)\mathrm{TOFF}, \\
    \mathrm{TOFF}(I \otimes Z \otimes I)\mathrm{TOFF},\ \mathrm{TOFF}(I \otimes X \otimes I)\mathrm{TOFF}, \\
    \mathrm{TOFF}(I \otimes I \otimes Z)\mathrm{TOFF}, \ \mathrm{TOFF}(I \otimes I \otimes X)\mathrm{TOFF}.
\end{align*}

Using Alg.~\ref{alg:correction_operator}, we obtain that the Pauli decompositions are, respectively
\begin{align}
    Z \otimes I \otimes I, \hspace{0.35\columnwidth} \label{circuit1} \\ 
    I \otimes Z \otimes I, \hspace{0.35\columnwidth} \label{circuit2} \\
    \resizebox{.9\linewidth}{!}{$\frac{1}{2}(I \otimes I \otimes Z + I \otimes Z \otimes Z + Z \otimes I \otimes Z - Z \otimes Z \otimes Z)$}, \label{circuit3} \\
    \resizebox{.9\linewidth}{!}{$\frac{1}{2}(X \otimes I \otimes I + X \otimes I \otimes X + X \otimes Z \otimes I - X \otimes Z \otimes X)$}, \label{circuit4} \\
    \resizebox{.9\linewidth}{!}{$\frac{1}{2}(I \otimes X \otimes I + I \otimes X \otimes X + Z \otimes X \otimes I - Z \otimes X \otimes X)$}, \label{circuit5} \\
    I \otimes I \otimes X. \hspace{0.35\columnwidth} \label{circuit6}
\end{align}

Every Pauli decomposition defines a Hamiltonian $e^{-iH\Delta t}$, whose time evolution can generally be approximated using Hamiltonian simulation techniques such as Trotterization. In this case, we note that we can apply the idea presented in \cite{Nielsen_Chuang_2010} and build simple circuits that implement $e^{-iH\Delta t}$ for arbitrary values of $\Delta t$. Depending on the results from the Bell measurements, we apply these Hamiltonians as correction operations.

Since we are interested in implementing the Hamiltonians in SquidASM, we construct their corresponding circuits here. Since the standard $Z$-gate corresponds to $\pi$ rotation with respect to $Z$-axis, we obtain that the Pauli decompositions having $\frac{1}{2}$ coefficient are rotated $\pi/2$ and those with the negative sign are rotated $-\pi/2$. Using this information and the idea presented in \cite{Nielsen_Chuang_2010}, the Pauli decompositions are easy to write as circuits.

\begin{figure}[tbp]
    \centering
    \resizebox{\columnwidth}{!}{
    \input{circuits/toffoli_circuits/circuit4}
    }
    \caption{Pauli decomposition \eqref{circuit4}}
    \label{fig:circuit4}
\end{figure}
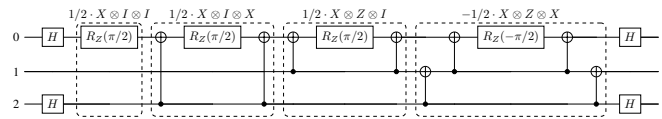

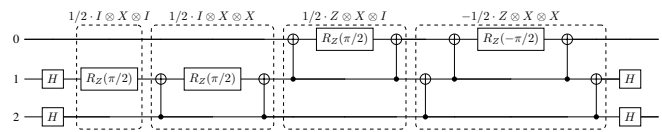
\begin{figure}[tbp]
    \centering
    \resizebox{\columnwidth}{!}{
    \input{circuits/toffoli_circuits/circuit5}
    }
    \caption{Pauli decomposition \eqref{circuit5}}
    \label{fig:circuit5}
\end{figure}

Fig.~\ref{fig:circuit1} represents \eqref{circuit1}, Fig.~\ref{fig:circuit2} represents \eqref{circuit2}, Fig.~\ref{fig:circuit3} represents \eqref{circuit3}, Fig.~\ref{fig:circuit4} represents \eqref{circuit4}, Fig.~\ref{fig:circuit5} represents \eqref{circuit5}, and Fig.~\ref{fig:circuit6} represents \eqref{circuit6}.

\begin{figure}[tbp]
\centering
\begin{subfigure}[t]{.25\columnwidth}
    \centering
    \resizebox{\columnwidth}{!}{%
    \input{circuits/toffoli_circuits/circuit1}
    }
    \caption{Pauli decomposition~\eqref{circuit1}}
    \label{fig:circuit1}
\end{subfigure}%
\begin{subfigure}[t]{.25\columnwidth}
    \centering
    \resizebox{\columnwidth}{!}{%
    \input{circuits/toffoli_circuits/circuit2}
    }
    \caption{Pauli decomposition~\eqref{circuit2}}
    \label{fig:circuit2}
\end{subfigure}%
\begin{subfigure}[t]{.39\columnwidth}
    \centering
    \resizebox{\columnwidth}{!}{%
    \input{circuits/toffoli_circuits/circuit6}
    }
    \caption{Pauli decomposition ~\eqref{circuit6}}
    \label{fig:circuit6}
\end{subfigure}
\caption{Simple circuits implementing Pauli decompositions}
\label{fig:simple_circuits}
\end{figure}
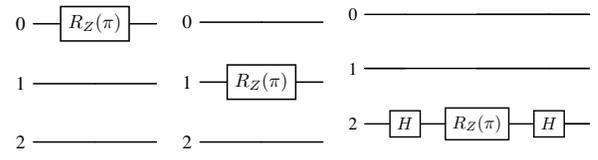

\begin{figure}[tbp]
    \centering
    \resizebox{\columnwidth}{!}{%
    \input{circuits/toffoli_circuits/circuit3}
    }
    \caption{The circuit implementing Pauli decomposition \eqref{circuit3}}
    \label{fig:circuit3}
\end{figure}
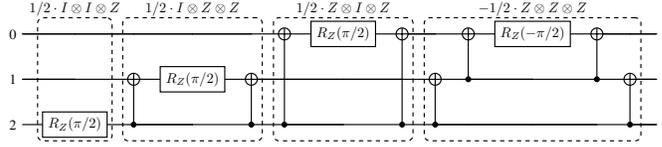

Finally, we obtain the full Toffoli teleportation circuit represented in Fig.~\ref{fig:full_toffoli_teleportation}. We have implemented Toffoli gate teleportation with two different quantum computing frameworks: Pennylane and SquidASM. Pennylane implementation is represented in Fig.~\ref{fig:toffoli_in_pennylane}. The implementation uses mid-circuit measurements and operations controlled by the measurement results. We have created six Pennylane functions implementing the circuit operations in Fig.~\ref{fig:circuit1}--\ref{fig:circuit5}. The code to run the Pennylane circuit can be found on GitHub \cite{SquidKnit}.

\begin{figure*}[tbp]
    \centering
    \resizebox{\textwidth}{!}{%
     \input{circuits/toffoli_circuits/full_toffoli_teleportation}
    }
    \caption{Circuit implementing Toffoli teleportation. Depending on the six bits we measure, different circuits are applied to the last three output qubits. The numbering in the correction operations refers to Pauli decompositions \eqref{circuit1} -- \eqref{circuit6}. The circuits implementing these decompositions are in Fig. \ref{fig:circuit1} -- \ref{fig:circuit5}.}
    \label{fig:full_toffoli_teleportation}
\end{figure*}
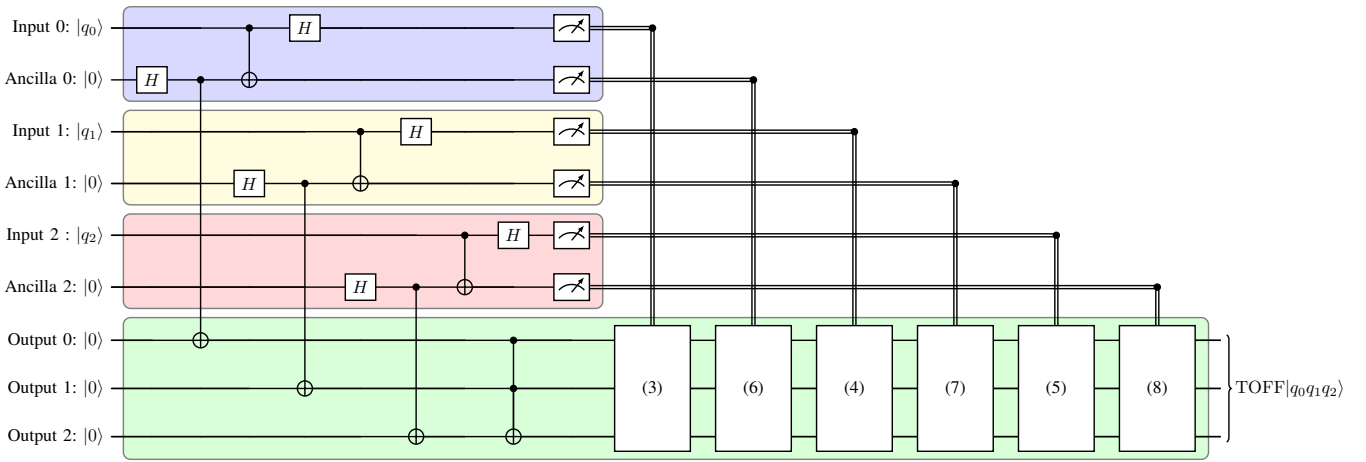

\begin{figure}[tbp]
    \centering
    \begin{lstlisting}[language=Python]
@qml.qnode(quantum_device)
def toffoli_teleportation(rots):
    # 0,1,2: Input qubits
    # 3,4,5: Ancilla qubits
    # 5,6,7: Output qubits

    # Prepare the random input state for Toffoli 
    # gate using some random rotations
    qml.Rot(rots[0], rots[1], rots[2], wires=0)
    qml.Rot(rots[3], rots[4], rots[5], wires=1)
    qml.Rot(rots[6], rots[7], rots[8], wires=2)
    
    # Prepare the Bell pairs
    bell_pair(3, 6)
    bell_pair(4, 7)
    bell_pair(5, 8)
    
    #Prepare Bell measurement
    bell_measurement(0, 3)
    bell_measurement(1, 4)
    bell_measurement(2, 5)
    
    # Perform the Toffoli gate on the output qubits
    qml.Toffoli(wires=[6, 7, 8])
    
    # Perform the Bell measurement as 
    # mid-circuit measurements
    m0 = qml.measure([0], reset=True)
    m1 = qml.measure([1], reset=True)
    m2 = qml.measure([2], reset=True)
    a0 = qml.measure([3], reset=True)
    a1 = qml.measure([4], reset=True)
    a2 = qml.measure([5], reset=True)

    # Perform circuits conditioned on 
    # mid-circuit measurements
    coeff = np.pi/2
    qml.cond(m0, circuit_2)(coeff=2*coeff)
    qml.cond(m1, circuit_3)(coeff=2*coeff)
    qml.cond(m2, circuit_4)(coeff=coeff)
    qml.cond(a0, circuit_5)(coeff=coeff)
    qml.cond(a1, circuit_6)(coeff=coeff)
    qml.cond(a2, circuit_7)(coeff=2*coeff)
    
    return qml.density_matrix(wires=[6, 7, 8])
    \end{lstlisting}
    \caption{Circuit implementing Toffoli teleportation with Pennylane using the circuits in Fig. \ref{fig:circuit1}--\ref{fig:circuit5}. The circuit numbering, i.e. \texttt{circuit\_n} for \texttt{n} $= 2, \ldots, 7$, in the code matches the numbers in Fig. \ref{fig:full_toffoli_teleportation}.}
    \label{fig:toffoli_in_pennylane}
\end{figure}

We have also implemented Toffoli teleportation using SquidASM. The implementation follows the previously presented implementations for two and three-node simulations. The division of the Toffoli teleportation circuit is divided with respect to the coloring in Fig.~\ref{fig:full_toffoli_teleportation}. This implementation creates a four-node quantum network where we have three InputPrograms (Alg.~\ref{alg:input_program}) and one GateProgram (Alg.~\ref{alg:gate_program}). As far as we know, this is one of the first four-node quantum network applications and one of the first concrete demonstrations of teleporting a non-Clifford gate.

%% file: circuits/toffoli_circuits/circuit1.tex
\begin{quantikz}
\lstick{0} & \gate{R_Z(\pi)} & \\
\lstick{1} & \ghost{R_Z} & \\
\lstick{2} & \ghost{R_Z} &
\end{quantikz}

%% file: circuits/toffoli_circuits/circuit2.tex
\begin{quantikz}
\lstick{0} & \ghost{R_Z} & \\
\lstick{1} & \gate{R_Z(\pi)} & \\
\lstick{2} & \ghost{R_Z} &
\end{quantikz}

%% file: circuits/toffoli_circuits/circuit6.tex
\begin{quantikz}
\lstick{0} & \ghost{R_Z} & \ghost{R_Z} & \qw & \qw  \\
\lstick{1} & \qw & \ghost{R_Z} & \qw & \qw \\
\lstick{2} & \gate{H} & \gate{R_Z(\pi)} & \gate{H} & \qw \\
\end{quantikz}

%% file: circuits/toffoli_circuits/circuit3.tex
\begin{quantikz}
\lstick{0} & \qw \gategroup[3,steps=1,style={dashed, rounded corners, inner ysep=-1pt, inner xsep=0pt}]{$1/2 \cdot I \otimes I \otimes Z$} & \qw \gategroup[3,steps=3,style={dashed, rounded corners, inner ysep=-1pt, inner xsep=0pt}]{$1/2 \cdot I \otimes Z \otimes Z$} & \qw & \qw & \targ{} \gategroup[3,steps=3,style={dashed, rounded corners, inner ysep=-1pt, inner xsep=0pt}]{$1/2 \cdot Z \otimes I \otimes Z$} & \gate{R_Z(\pi/2)} & \targ{} & \qw \gategroup[3,steps=5,style={dashed, rounded corners, inner ysep=-1pt, inner xsep=0pt}]{$-1/2 \cdot Z \otimes Z \otimes Z$} & \targ{} & \gate{R_Z(-\pi/2)} & \targ{} & \qw & \qw \\
\lstick{1} & \qw & \targ{} & \gate{R_Z(\pi/2)} & \targ{} & \qw & \qw & \qw & \targ{} & \ctrl{-1} & \qw & \ctrl{-1} & \targ{} & \qw \\
\lstick{2} & \gate{R_Z(\pi/2)} & \ctrl{-1} & \qw & \ctrl{-1} & \ctrl{-2} & \qw & \ctrl{-2} & \ctrl{-1} & \qw & \qw& \qw & \ctrl{-1} & \qw
\end{quantikz}

%% file: circuits/toffoli_circuits/circuit4.tex
\begin{quantikz}
\lstick{0} & \gate{H} & \gate{R_Z(\pi/2)}\gategroup[3,steps=1,style={dashed, rounded corners, inner ysep=-1pt, inner xsep=0pt}]{$1/2 \cdot X \otimes I \otimes I$} & \targ{} \gategroup[3,steps=3,style={dashed, rounded corners, inner ysep=-1pt, inner xsep=0pt}]{$1/2 \cdot X \otimes I \otimes X$} & \gate{R_Z(\pi/2)} & \targ{} & \targ{} \gategroup[3,steps=3,style={dashed, rounded corners, inner ysep=-1pt, inner xsep=0pt}]{$1/2 \cdot X \otimes Z \otimes I$} & \gate{R_Z(\pi/2)} & \targ{} & \qw \gategroup[3,steps=5,style={dashed, rounded corners, inner ysep=-1pt, inner xsep=0pt}]{$-1/2 \cdot X \otimes Z \otimes X$} & \targ{} & \gate{R_Z(-\pi/2)} & \targ{} & \qw & \gate{H} & \qw \\
\lstick{1} & \qw & \qw & \qw & \qw & \qw & \ctrl{-1} & \qw & \ctrl{-1} & \targ{} & \ctrl{-1} & \qw & \ctrl{-1} & \targ{} & \qw & \qw \\
\lstick{2} & \gate{H} & \qw & \ctrl{-2} & \qw & \ctrl{-2} & \qw & \qw & \qw & \ctrl{-1} & \qw & \qw & \qw & \ctrl{-1} & \gate{H} & \qw
\end{quantikz}

%% file: circuits/toffoli_circuits/circuit5.tex
\begin{quantikz}
\lstick{0} & \qw & \qw \gategroup[3,steps=1,style={dashed, rounded corners, inner ysep=-1pt, inner xsep=0pt}]{$1/2 \cdot I \otimes X \otimes I$} & \qw \gategroup[3,steps=3,style={dashed, rounded corners, inner ysep=-1pt, inner xsep=0pt}]{$1/2 \cdot I \otimes X \otimes X$} & \qw & \qw & \targ{} \gategroup[3,steps=3,style={dashed, rounded corners, inner ysep=-1pt, inner xsep=0pt}]{$1/2 \cdot Z \otimes X \otimes I$} & \gate{R_Z(\pi/2)} & \targ{} & \qw \gategroup[3,steps=5,style={dashed, rounded corners, inner ysep=-1pt, inner xsep=0pt}]{$-1/2 \cdot Z \otimes X \otimes X$} & \targ{} & \gate{R_Z(-\pi/2)} & \targ{} & \qw & \qw & \qw \\
\lstick{1} & \gate{H} & \gate{R_Z(\pi/2)} & \targ{} & \gate{R_Z(\pi/2)} & \targ{} & \ctrl{-1} & \qw & \ctrl{-1} & \targ{} & \ctrl{-1} & \qw & \ctrl{-1} & \targ{} & \gate{H} \qw \\
\lstick{2} & \gate{H} & \qw & \ctrl{-1} & \qw & \ctrl{-1} & \qw & \qw & \qw & \ctrl{-1} & \qw & \qw & \qw & \ctrl{-1} & \gate{H} & \qw
\end{quantikz}

%% file: circuits/toffoli_circuits/full_toffoli_teleportation.tex
\begin{quantikz}
\lstick{Input 0: \ket{q_0}} & \qw \gategroup[2,steps=9,style={rounded corners,fill=blue!30,opacity=0.5,inner ysep=1pt}, background]{} & \qw & \ctrl{1} & \gate{H} & \qw & \qw & \qw & \qw & \meter{} & \ctrl[vertical wire=c]{6} \setwiretype{c} \\
\lstick{Ancilla 0: \ket{0}} & \gate{H} & \ctrl{5} & \targ{} & \qw & \qw & \qw & \qw & \qw & \meter{} & \setwiretype{c} & \ctrl[vertical wire=c]{5} \\
\lstick{Input 1: \ket{q_1}} & \qw \gategroup[2,steps=9,style={rounded corners,fill=yellow!30,opacity=0.5,inner ysep=1pt}, background]{} & \qw & \qw & \qw & \ctrl{1} & \gate{H} & \qw & \qw & \meter{} & \setwiretype{c} & &\ctrl[vertical wire=c]{4} \\
\lstick{Ancilla 1: \ket{0}} & \qw & \qw &  \gate{H} & \ctrl{4} & \targ{} & \qw & \qw & \qw & \meter{} & \setwiretype{c} & & &\ctrl[vertical wire=c]{3} \\
\lstick{Input 2 : \ket{q_2}} & \qw \gategroup[2,steps=9,style={rounded corners,fill=red!30,opacity=0.5,inner ysep=1pt}, background]{} & \qw & \qw & \qw & \qw & \qw & \ctrl{1} & \gate{H} & \meter{} & \setwiretype{c} & & & &\ctrl[vertical wire=c]{2} \\
\lstick{Ancilla 2: \ket{0}} & \qw & \qw &  \qw & \qw & \gate{H} & \ctrl{3} & \targ{} & \qw & \meter{} & \setwiretype{c} & & & & &\ctrl[vertical wire=c]{1} \\
\lstick{Output 0: \ket{0}} & \qw \gategroup[3,steps=15,style={rounded corners,fill=green!30,opacity=0.5,inner ysep=1pt}, background]{} & \targ{} & \qw & \qw & \qw & \qw & \qw & \ctrl{2} & \qw & \gate[3][1.5cm]{\eqref{circuit1}} & \gate[3][1.5cm]{\eqref{circuit4}} & \gate[3][1.5cm]{\eqref{circuit2}} & \gate[3][1.5cm]{\eqref{circuit5}} & \gate[3][1.5cm]{\eqref{circuit3}} & \gate[3][1.5cm]{\eqref{circuit6}} & \qw \rstick[3]{$ \mathrm{TOFF}|q_0q_1q_2\rangle$} \\
\lstick{Output 1: \ket{0}} & \qw & \qw & \qw & \targ{} & \qw & \qw & \qw & \ctrl{1} & \qw & \qw & \qw & \qw & \qw & \qw & \qw & \qw \\
\lstick{Output 2: \ket{0}} & \qw & \qw & \qw & \qw & \qw & \targ{} & \qw & \targ{} & \qw & \qw & \qw & \qw & \qw & \qw & \qw & \qw
\end{quantikz}

%% file: discussion.tex
\subsection{Results}

Considering Clifford gate teleportation in two- and three-node quantum networks, we performed the gate teleportation for CNOT, DCNOT, CZ, and SWAP gates. Before teleportation, the user can prepare the input state as their application requires. We decided to apply Hadamard gates, i.e., $|q_0\rangle = |q_1\rangle = H|0\rangle$ where $|q_0\rangle$ and $|q_1\rangle$ are the input states for the gate $U$. We performed 100 runs for each simulation, calculated the fidelity with respect to the exact state $U |q_0 q_1 \rangle$ computed without noise, and averaged the results.

\begin{figure}[tbp]
    \centering
    \includegraphics[width=\columnwidth]{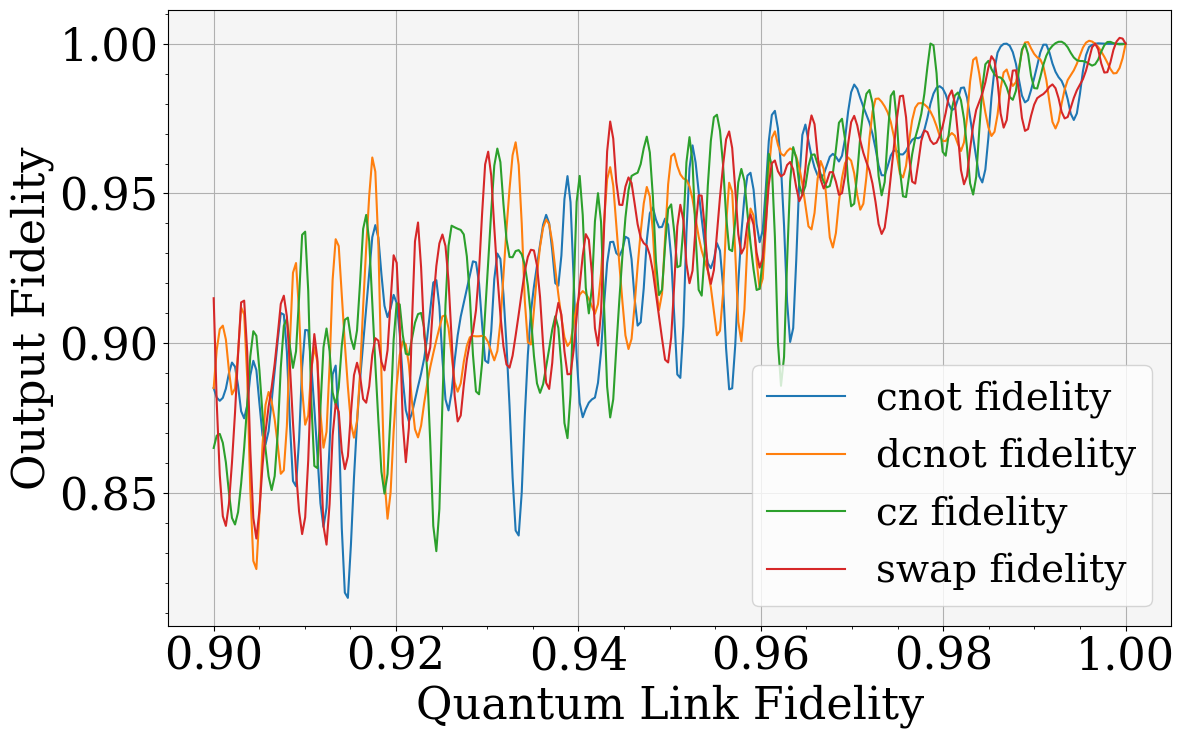}
    \caption{Fidelity of results when quantum channel is noisy}
    \label{fig:link_fidelity_2}
\end{figure}

\begin{figure}[tbp]
    \centering
    \includegraphics[width=\columnwidth]{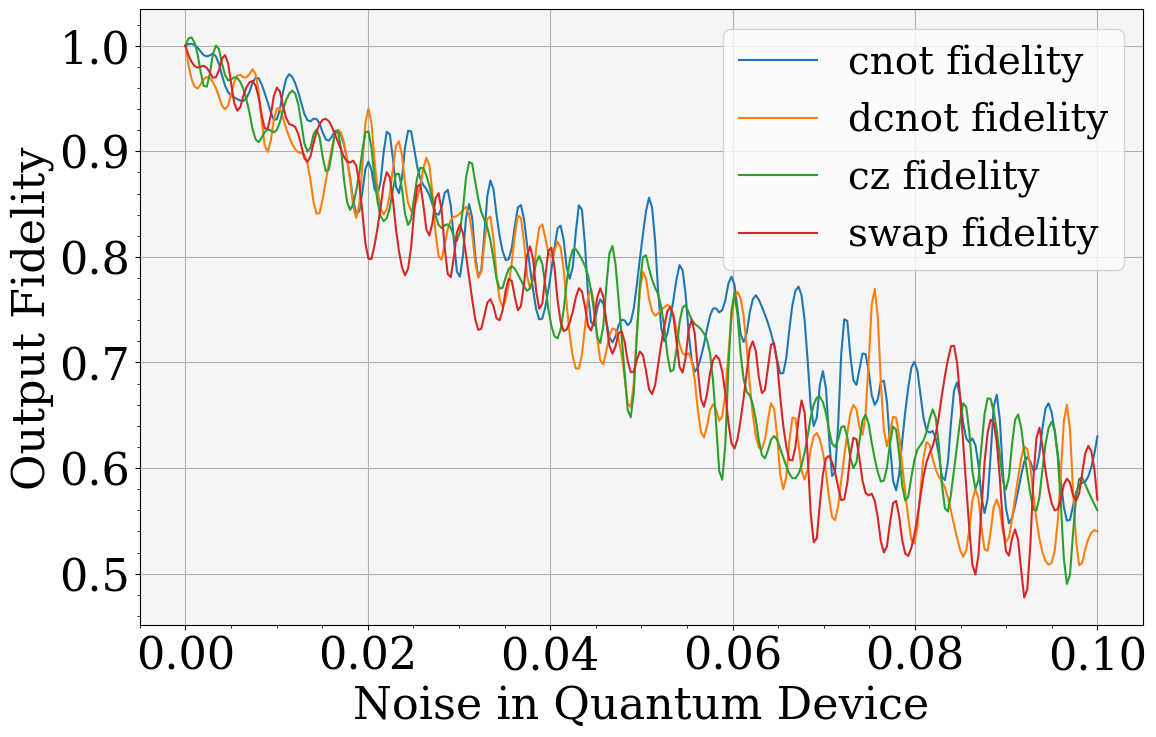}
    \caption{Fidelity of results when quantum devices are noisy}
    \label{fig:qdevice_fidelity_2}
\end{figure}

Fig. \ref{fig:link_fidelity_2} represents the results from the simulation in the case that we varied the quantum channel's fidelity in the two-node network. We initialized the link with the simple depolarised link configuration. We used SquidASM's Generic Quantum Device to perform noiseless simulations on the quantum device side so that only noise appeared in the quantum channel. The classical channel used the default configuration. In the other simulation, we used a similar setup but varied the noise in the quantum device. These results are in Fig. \ref{fig:qdevice_fidelity_2}.


\begin{figure}[tbp]
    \centering
    \includegraphics[width = 0.99\columnwidth]{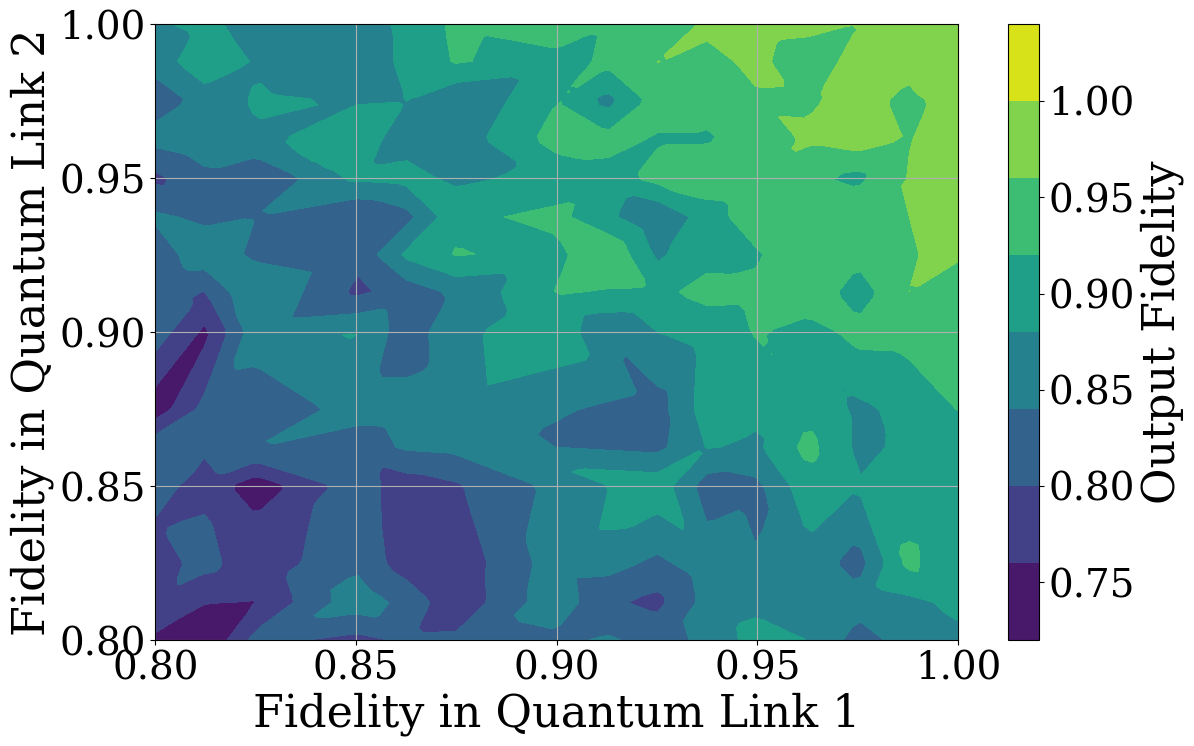}
    \caption{Output fidelity when varying two quantum channels}
    \label{fig:three_node_link_evaluation}
\end{figure}

\begin{figure}[tbp]
    \centering
    \includegraphics[width = 0.99\columnwidth]{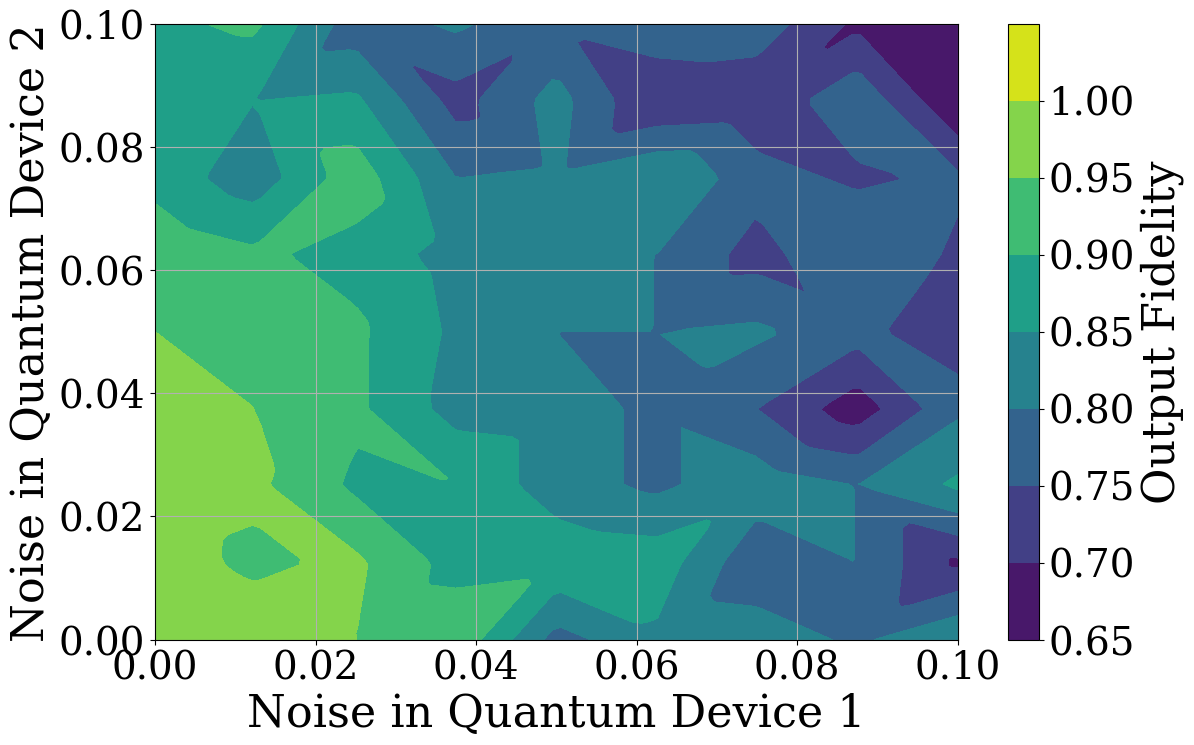}
    \caption{Output fidelity when varying noise on two quantum devices}
    \label{fig:qdevice_noise_three_nodes}
\end{figure}

In the three-node quantum network simulation, we used two quantum channels. We have performed experiments including varying noise in both quantum channels (Fig.~\ref{fig:three_node_link_evaluation}). Then, we performed simulations that introduced varying noise to one of the InputPrograms and the GateProgram (Fig.~\ref{fig:qdevice_noise_three_nodes}).

\begin{figure}[tbp]
    \centering
    \includegraphics[width = 0.99\columnwidth]{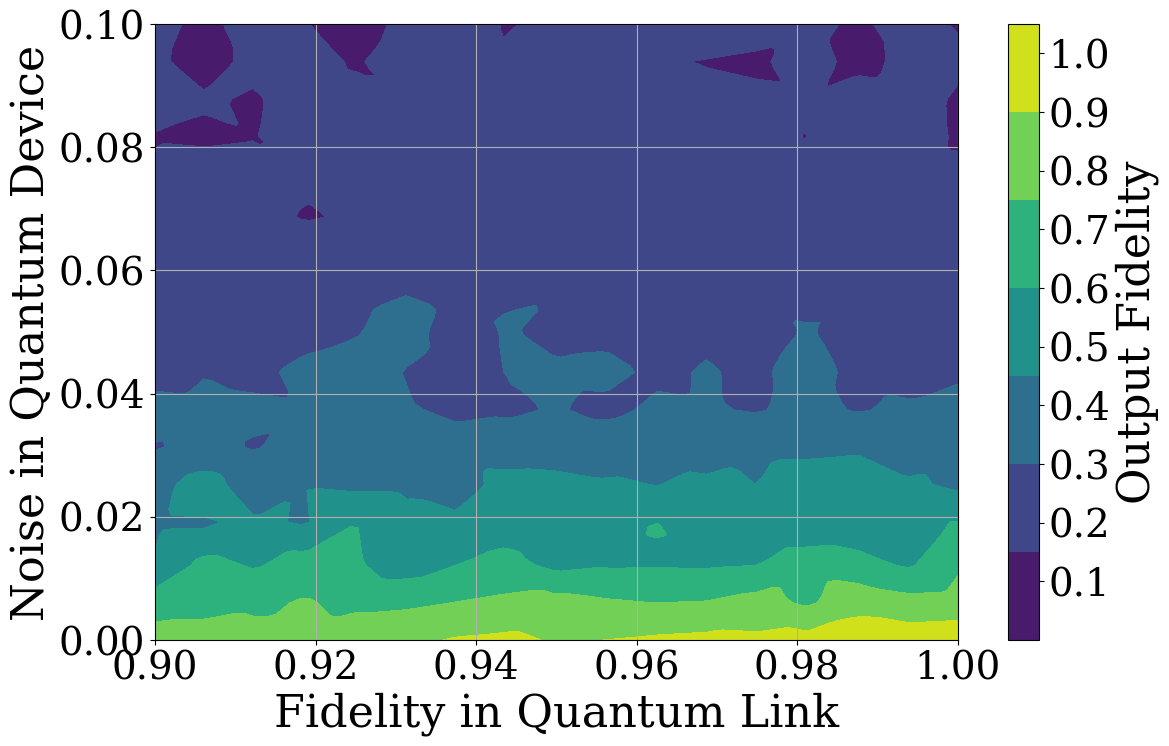}
    \caption{Results from teleporting Toffoli gate}
    \label{fig:toffoli_results}
\end{figure}

Finally, we simulated the four-node quantum network teleporting the Toffoli gate. In this case, we decided to vary both channel and quantum device noise simultaneously. More precisely, we set the InputPrograms (blue, yellow, and red machines in Fig.~\ref{fig:full_toffoli_teleportation}) to have zero noise so that the quantum device noise appears only in GateProgram where we perform the correction operation. This is interesting because then we can discover how much the relatively long correction operation affects the computation. Simultaneously, we varied the noise in the three different quantum channels. These results are presented in Fig.~\ref{fig:toffoli_results}.

\section{Discussion}
Comparing the output fidelities of the two-node simulations depicted in Fig.~\ref{fig:link_fidelity_2} and Fig.~\ref{fig:qdevice_fidelity_2} or the output fidelities of the three-node simulations in Fig.~\ref{fig:three_node_link_evaluation} and Fig.~\ref{fig:qdevice_noise_three_nodes}, we observe that noise affecting the link has a lesser impact on computation output compared to noise affecting the device. This is expected since the gate operations happen on the devices and are more exposed to the noise. At least in the context of these experiments, the findings do not rule out the idea of connecting multiple high-performing but small quantum computers with quantum links that have a small error. In this sense, the results support the idea that small but high-quality quantum computers should be connected with ''satisfying'' links rather than building equal-sized but more error-prone quantum computers.

The results from simulating the Toffoli gate teleportation \ref{fig:toffoli_results} show a similar pattern even more clearly. This happened because the correction operation is the longest operation in the Toffoli gate teleportation, and it is substantially longer than the correction operations in non-Clifford gate teleportation schemes. All in all, the simulations confirm that SquidASM is capable of simulating noisy multi-node quantum networks, and noise in the quantum device affects computation more than the noise in the quantum channel.

Besides that quantum gate teleportation can be used to realize the gate-cutting method (Fig.~\ref{fig:diagram}), its original motivation is to implement fault-tolerant quantum computing \cite{Gottesman_1998, Gottesman_Chuang_1999, Gottesman_Chuang_1999_universal}. Intuitively, the gate can be performed on a receiver's device so that all the operations before the gate are simple (CNOT and Hadamard), as Fig.~\ref{fig:single_gate_teleportation} demonstrates. For a fixed gate, we are always preparing the same known state, which is easier than operating the gate on an unknown state \cite{Gottesman_Chuang_1999}. The presumably complicated input state for the gate is not lost even if the preparation of the gate fails on the receiver's device. 

Now, the gate teleportation protocol can be used to perform fault-tolerant quantum computing with a recursive scheme that is based on set \eqref{eq:clifford_k}. Let $U \in C_k$ be a gate we want to teleport. If we are able to prepare the $U$ on the receiver's side, then the correction operation is an element of $C_{k-1}$ by the definition of set \eqref{eq:clifford_k}. Assuming that we are able to initialize the gate $U$ for the simple state and the gates in $C_{k-1}$ fault-tolerantly, we can apply the gate $U$ fault-tolerantly to an arbitrary input state. This introduces a recursive scheme to perform any gate fault-tolerantly.

Of course, there are additional complexities, such as quantum networking usually involves a certain overhead in the number of qubits and requires operations that might be challenging to realize on real hardware. In order to realize the algorithms in real-life quantum computing ecosystems, crucial hardware developments are needed. Especially challenging is to implement mid-circuit measurements (in this work, Bell measurements) and be able to maintain the quantum states sufficiently long.

The three-node results and the theoretical background in Fig.~\ref{fig:diagram} demonstrate how gate cutting can be performed precisely without classical computing relying on a third quantum device. The results obtained from the three- and four-node simulations show that gate-cutting without classical computations is a possible approach to implementing large circuits in distributed quantum computing. Especially promising the result is because we are not restricted to Clifford gates but the gate teleportation protocol works for arbitrary gates. In the case of Toffoli gate, we also gain some understanding regarding the overhead that the correction method produces. The circuits in Fig.~\ref{fig:circuit1}--Fig.~\ref{circuit5} show that in the worst case, we need to perform about 50 gates to correct the result, which is substantially more than in the case of teleporting Clifford gates where the worst case is around two times the number of used qubits, i.e., three-qubit Clifford gate would require only six single qubit correcting gates. We can also see how the longer correction method decreases the quality of the Toffoli gate teleportation compared to the results obtained from the two-node networks (Fig.\ref{fig:qdevice_fidelity_2}) and three-node networks (Fig.~\ref{fig:qdevice_noise_three_nodes}).

As we noted in the beginning, quantum teleportation has mainly focused on hardware development. Quantum networks and circuit cutting bring new challenges to circuit compilation. Instead of optimizing circuit compilation for a single quantum computing topology, we have a collection of possibly varying topologies that are connected with quantum channels. If we are employing the circuit knitting technique, we should cut the circuits in a way that the classical and quantum overheads are minimized \cite{Piveteau_Sutter_2024}. A similar quantum compilation-related challenge persists if we perform gate teleportation-based circuit cutting.

Our paper and implementation are among the first multi-node quantum internet applications utilizing SquidASM and one of the first to use realistic noise models for a practical quantum network application. We did not limit our work to Clifford gates but also considered the Toffoli gate. Besides, we reviewed the motivated use cases of how the gate teleportation protocol can be utilized in gate cutting and fault-tolerant quantum computing. The presented concrete implementations are important in the future development of quantum networks in order to realize their practical utility.

\section{Conclusion and future work}

In this work, we started by reviewing the standard state and gate teleportation techniques. We implemented the algorithm to calculate the correction operator for arbitrary two-qubit Clifford gates. We developed two- and three-node quantum network simulations to perform two-qubit Clifford gate teleportation using the SquidASM quantum network simulator. We then developed the protocol to perform Toffoli gate teleportation and realized that with Pennylane and SquidASM. As an important application, these simulations demonstrated a gate-cutting method without additional classical overhead but using a third quantum device. We obtained various simulation results, which indicate that quantum channels might be a promising method for building larger noise-resilient quantum computers.

In future work, we are especially interested in developing and finding applications for quantum networks. One application field is quantum machine learning, where quantum networks and quantum teleportation-based schemes could be used in novel ways to train and run models in distributed environments. We believe we have not yet seen all the possible use cases of quantum gate teleportation.